\documentclass[fleqn,12pt,twoside]{article}
\usepackage{espcrc1}
\usepackage{floatflt}
\usepackage{wrapfig}

\def\xssnsn{$\sigma(AuAu \!\rightarrow\! Au^*_{1n} Au^*_{1n} \rho^0) \!=\!{ 2.8 \!\pm\! 0.5\!\pm\! 0.7}$~mb}
\def\xsxnxn{$\sigma(AuAu \!\rightarrow\! Au^*_{xn} Au^*_{xn} \rho^0) \!=\! { 39.7  \!\pm\! 2.8  \!\pm\! 9.7}$~mb}
\def\xsnobrk{$\sigma(AuAu \!\rightarrow\! Au Au \rho^0) \!=\! { 370  \!\pm\! 170 \pm 80}$~mb}

\def\xstot{$\sigma(AuAu \!\rightarrow\! Au^{(*)} Au^{(*)} \rho^0) \!=\! { 460  \!\pm\! 220 \pm 110}$~mb}

\usepackage{graphicx}
\usepackage[figuresright]{rotating}

\newcommand{\AmS}{{\protect\the\textfont2
  A\kern-.1667em\lower.5ex\hbox{M}\kern-.125emS}}

\title{ Coherent Vector Meson Production in Ultra-Peripheral Heavy-Ion 
Collisions at STAR}
\author{F.~Meissner~\address[MCSD]{Lawrence Berkeley National Laboratory\\
	 One Cyclotron Rd., Berkeley, CA 94720}, for the STAR Collaboration
\footnote{For the full author list and acknowledgements, see Appendix
"Collaborations" of this volume.}
}
\begin{document}
\maketitle

\begin{abstract}
We report the first observation of coherent $\rho^0$
production ($AuAu \!\rightarrow \!AuAu \rho^0$) and  $\rho^0$ production
accompanied by mutual nuclear Coulomb excitation ($AuAu \!\rightarrow
\!Au^\star Au^\star \rho^0$), and the observation of  coherent $e^+e^-$ pair 
production ($AuAu \! \rightarrow \! Au^\star Au^\star e^+e^-$) in
ultra-peripheral relativistic heavy-ion collisions (UPCs). We give
transverse momentum, mass, and rapidity distributions.  The cross
sections for coherent $\rho^0$ production at
$\sqrt{s_{NN}}\!=\!130$ and $200$~GeV are in
agreement with theoretical predictions, which treat $\rho^0$
production and nuclear excitation as independent processes.
\end{abstract}
\vspace*{0.2cm}

In ultra-peripheral heavy ion collisions, photon exchange,
photon-photon or photon-nuclear interactions take place at impact
parameters $b$ larger than twice the nuclear radius $R_A$, where no
nucleon-nucleon collisions occur~\cite{baurrev}.  Examples are nuclear
Coulomb excitation, electron-positron pair and meson production, and
vector meson production. 
\begin{wrapfigure}[13]{r}{.5\textwidth}
\includegraphics[width=7.5cm,height=2.3cm,bbllx=0pt,bblly=0pt,bburx=650pt,bbury=180pt]{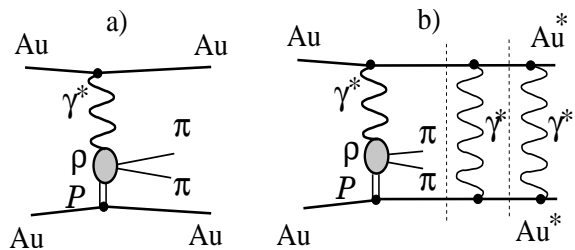}
\caption[]{ Diagram for (a) exclusive $\rho^0$ production in
ultra-peripheral heavy ion collisions, and (b) $\rho^0$ production
with  nuclear excitation. The dashed lines indicate factorization. 
\label{fig:feynman}}
\end{wrapfigure}
Exclusive $\rho^0$ meson production, $AuAu\!  \rightarrow\! Au
Au \rho^0$ (Fig.~\ref{fig:feynman}a), can be described by the
Weizs\"acker-Williams approach~\cite{weizsaecker} to the photon flux
and the vector meson dominance model~\cite{sakurai}.
A photon emitted by one nucleus fluctuates to a virtual $\rho^0$ meson, 
which scatters elastically from the other
nucleus. 
The gold nuclei are not disrupted, and the final state consists solely
of the two nuclei and the vector meson decay products~\cite{BKN}. In
addition to coherent $\rho^0$ production, the exchange of virtual
photons may excite the nuclei (Fig.~\ref{fig:feynman}b), yielding
the subsequent emission of single (1n) or multiple (xn) neutrons; these
processes are assumed to factorize for heavy-ion collisions.
In the rest frame of the target nucleus, 
mid-rapidity $\rho^0$ production at RHIC  corresponds to  a  photon energy of
$50$~GeV and a photon-nucleon center-of-mass energy of
$10$~GeV. At this energy, Pomeron $(\cal{P})$ exchange dominates over
meson exchange, as indicated by the rise of the $\rho^0$ production cross section with increasing 
energy in lepton-nucleon scattering.

The photon and Pomeron can couple coherently to the nuclei.
The wavelength $\lambda_{\gamma,\cal{P}}\!>\!2R_A$ leads
to coherence conditions: a low transverse momentum of $p_T \!<\! \pi
\hbar/ R_A$ ($\!\sim\!90$~MeV/c for gold with $R_A \!\sim\! 7$~fm),
and a maximum longitudinal momentum of $p_\|\!< \! \pi \hbar \gamma /
R_A$ ($\!\sim\!6(9)$~GeV/c at $\gamma\!=\!70(100)$), where $\gamma$ is the
Lorentz boost of the nucleus. The photon flux is
proportional to the square of the nuclear charge
$Z^2$~\cite{weizsaecker}, and the forward cross section for
elastic $\rho^0 A$ scattering  $d\sigma^{\rho A}/dt|_{t=0}$  scales as $A^{4/3}$ for surface
coupling and $A^2$ in the bulk limit.  
The $\rho^0$ production cross sections are large.
At a center-of-mass energy of
$\sqrt{s_{NN}}\!=\!130(200)$~GeV, a
total $\rho^0$ cross section, regardless of nuclear excitation,
$\sigma(AuAu
\!\rightarrow\!Au^{(\star)}Au^{(\star)}\rho^0)\!=\!350(590)$~mb is
predicted from a Glauber extrapolation of $\gamma p \!\rightarrow \! 
\rho^0 p$ data~\cite{BKN}.  Calculations for coherent $\rho^0$
production with nuclear excitation assume that both
processes are independent, sharing only a common impact
parameter~\cite{BKN,xsectAuAu}.

In the years 2000 and 2001, RHIC collided gold nuclei at a
center-of-mass energy of $\sqrt{s_{NN}}\!=\! 130$ and
$200$~GeV, respectively. The STAR detector consists of a 4.2~m
long cylindrical time projection chamber (TPC) of 2~m radius. In 2000(2001)
the TPC was operated in a 0.25(0.5)~T solenoidal magnetic field.  Particles
are identified by their energy loss in the TPC.  A central trigger
barrel (CTB) of $240$ scintillators surrounds the TPC. Two zero degree
calorimeters (ZDC) at $\pm$ 18m from the interaction point are
sensitive to the neutral remnants of nuclear break-up.

Exclusive $\rho^0$ production in UPC has a
distinctive experimental signature: the $\pi^+\pi^-$ decay products of
the $\rho^0$ meson are observed in an otherwise 'empty'
spectrometer. The tracks are approximately back-to-back in the
transverse plane due to the small $p_T$ of the pair.
Two different triggers are used for this analysis.  For $AuAu\! 
\rightarrow\! AuAu \rho^0$, about 30,000 (2000) and 1.5~M(2001) events
were collected using a low-multiplicity `topology' trigger.  The CTB
was divided in four azimuthal quadrants. Single hits were required in
the opposite side quadrants; the top and bottom quadrants acted as
vetoes to suppress cosmic rays.  
A fast on-line reconstruction removed events without reconstructible tracks
from the data stream.  To study $Au Au
\!\rightarrow\! Au^\star Au^\star \rho^0$, about 0.8~M(2000) and 2.5~M(2001) 
`minimum bias' events, which required coincident detection of neutrons
from nuclear break-up in both ZDCs as a trigger, are used for the
analysis.

\begin{figure}[!htb]
\vspace*{-1cm}
\begin{minipage}[t]{3.8cm}
\includegraphics[width=4.1cm,height=4.2cm]{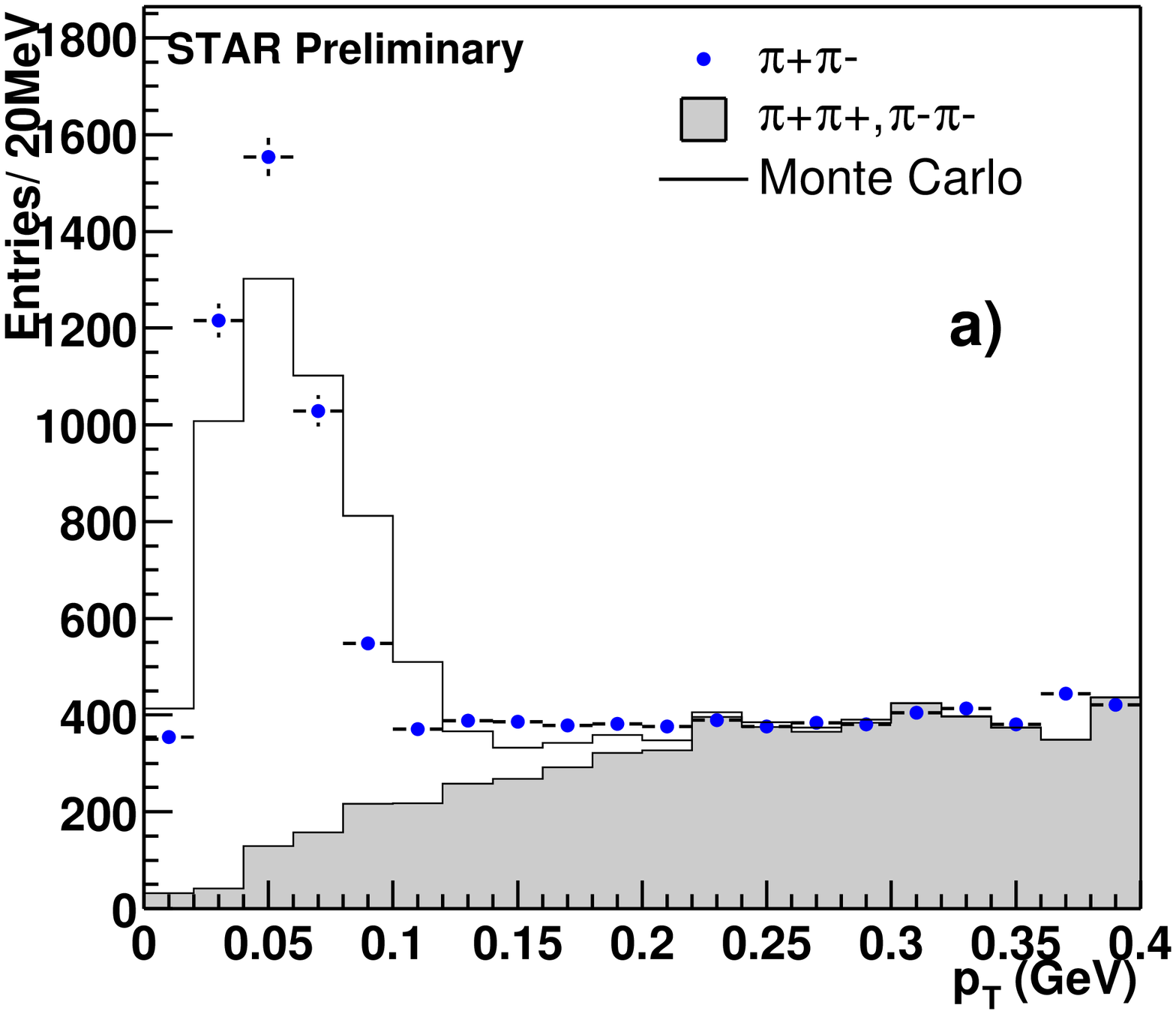}
\end{minipage}
\hspace{\fill}
\begin{minipage}[t]{3.8cm}
\includegraphics[width=4.1cm,height=4.2cm]{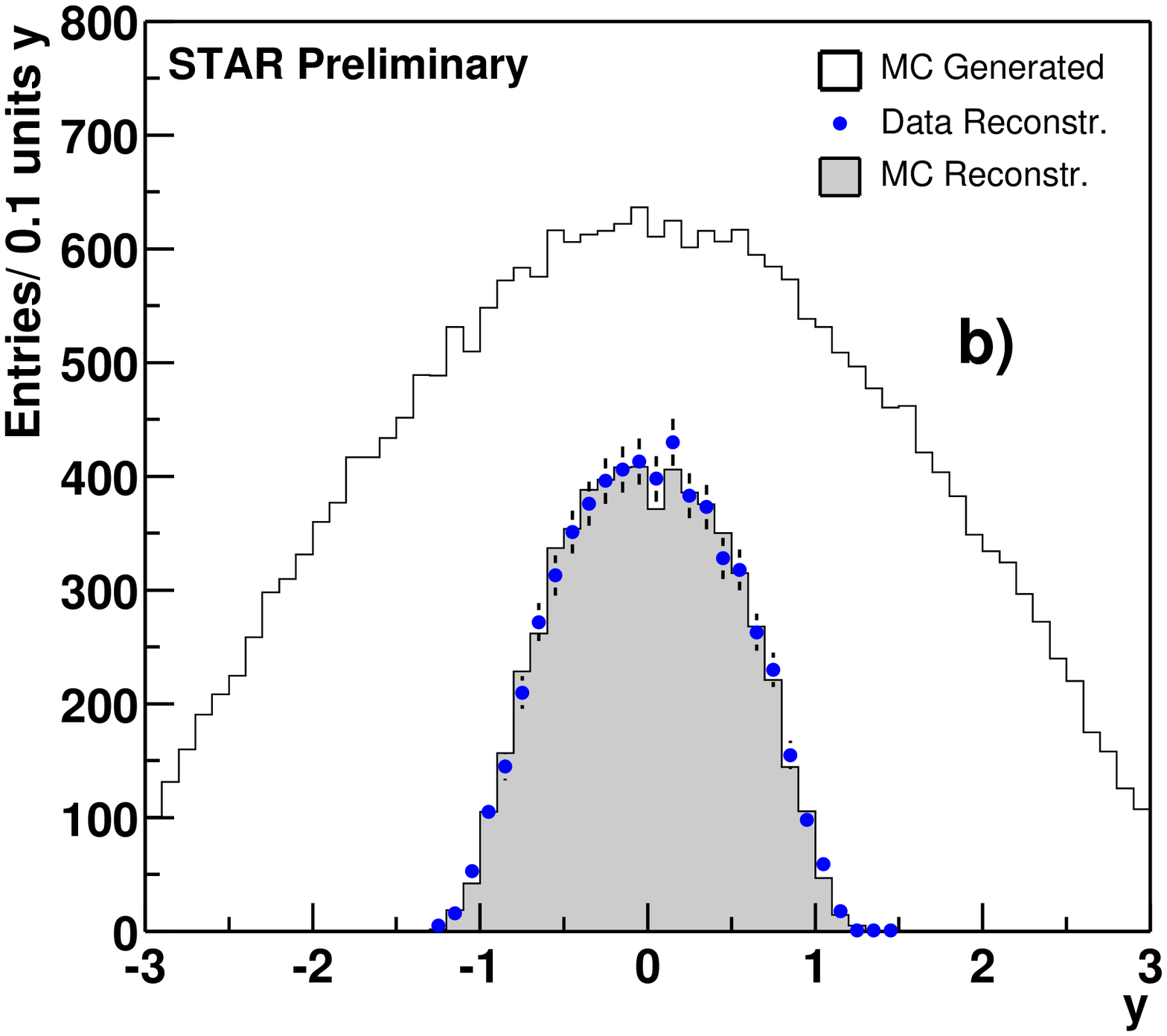}
\end{minipage}
\hspace{\fill}
\begin{minipage}[t]{3.8cm}
\includegraphics[width=4.1cm,height=4.2cm]{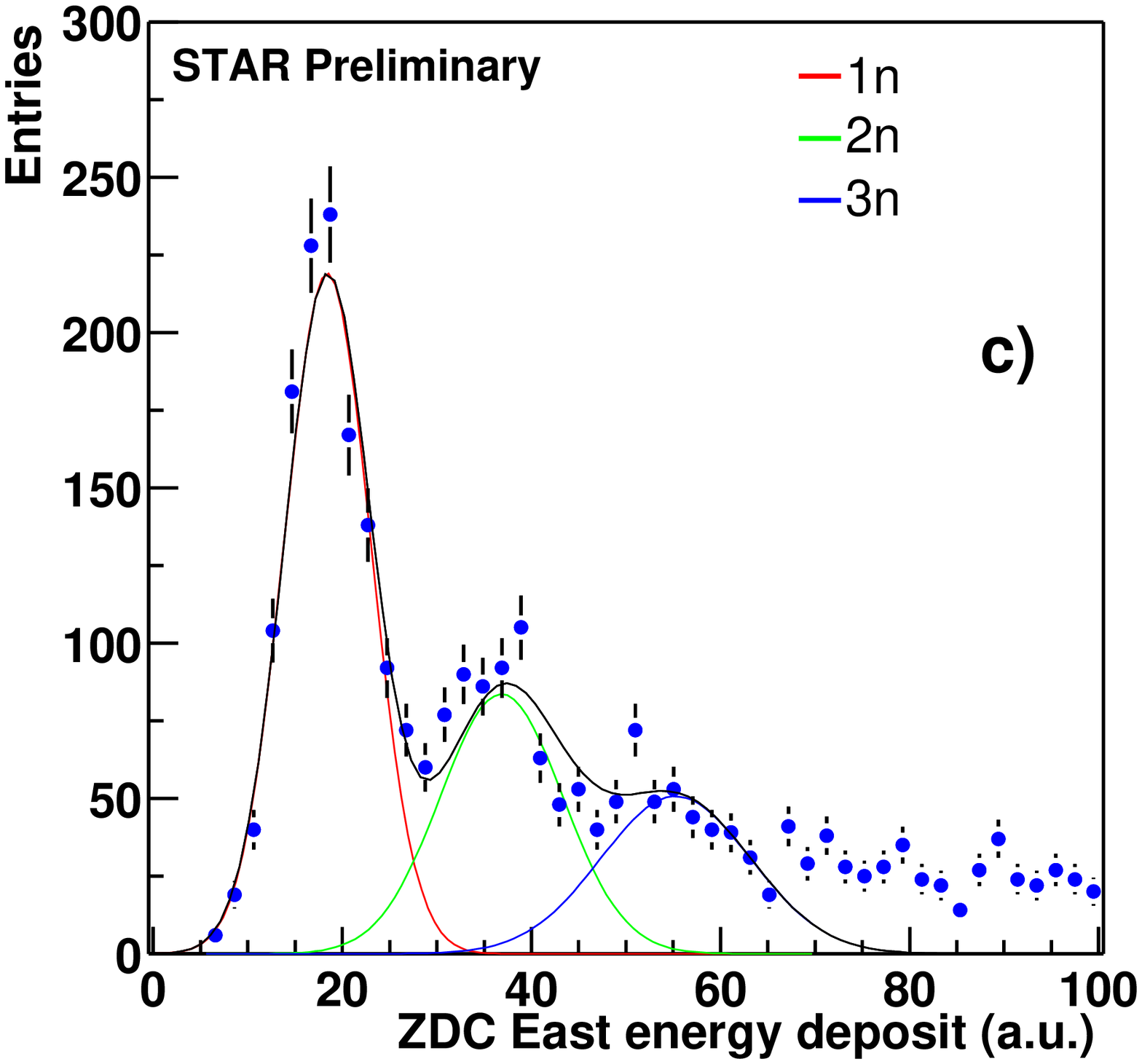}
\end{minipage}
\begin{minipage}[t]{3.8cm}
\includegraphics[width=4.1cm,height=4.2cm]{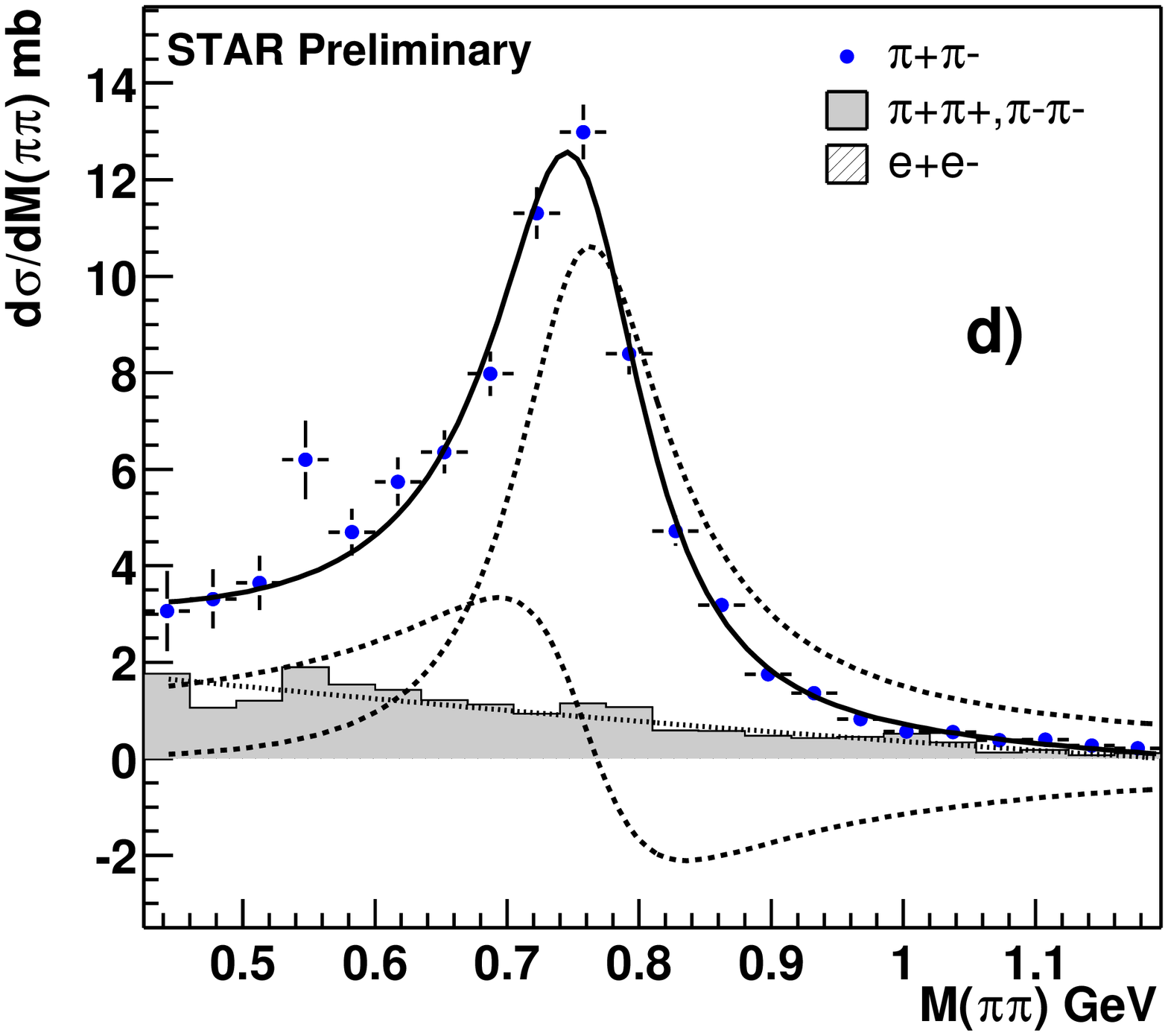}
\end{minipage}
\vspace*{-1cm}
\caption{
The (a) $\rho^0$ transverse momentum  and (b) rapidity distribution,
the (c) ZDC response, and (d) the $d\sigma/dM_{\pi\pi}$ invariant mass
distribution for  2-track (xn,xn) events in the $\sqrt{s_{NN}}\!=\!200$~GeV minimum bias data. 
\label{fig:all}}
\vspace*{-0.7cm}
\end{figure}

Events are selected with exactly two oppositely charged tracks forming
a common vertex within the interaction region.  The specific energy
loss $dE/dx$ in the TPC shows that the event sample is dominated by
pion pairs. In the topology triggered data sets, without the ZDC
requirement, cosmic rays are a major background.  They are removed by
requiring that the two pion tracks have an opening angle of less than
3 radians.
Figure~\ref{fig:all} shows kinematic distributions for the selected
2-track events in the $\sqrt{s_{NN}}\!=\!200$~GeV minimum bias data; these
distributions  are similar for the other data sets.

Figure~\ref{fig:all}a) shows the transverse momentum spectrum of
oppositely charged pion--pairs (points).  A clear peak, the signature
for coherent coupling, can be observed at $p_T\!<\!150$~MeV. Those
events are compatible with coherently produced $\rho^0$ candidates.  A
background model from like-sign combination pairs (shaded histogram),
which is normalized to the signal at $ p_T \!>\!$ 250 MeV, does not
show such a peak.  The open histogram is a Monte Carlo
simulation~\cite{BKN} for coherent $\rho^0$ production accompanied by
nuclear break-up superimposed onto the background.  The $dN^\rho/dp_T$
(i.e. the $dN^\rho/dt \!\sim \!dN^\rho/dp_T^2$) spectrum reflects not
only the nuclear form factor, but also the photon $p_T$ distribution
and the interference of production amplitudes from both gold
nuclei. The interference arises since both nuclei can be either the
photon source or the scattering target~\cite{vminterf}.  A
detailed analysis of the $p_T(t)$ distribution is in progress.

The rapidity distribution in Fig.~\ref{fig:all}b) is well
described by the reconstructed events from the Monte Carlo simulation.
The generated rapidity distribution is also shown.  The acceptance
for exclusive $\rho^0$ is about $40\%$ at $|y_\rho|\!<\!1$.
At $|y_\rho|\!>\!1$, the acceptance is small and this region is excluded from the analysis; the cross
sections are extrapolated to the full $4\pi$ acceptance with the Monte Carlo simulation.
Using the energy deposits in the ZDCs (Fig.~\ref{fig:all}c)), we select
events with at least one neutron (xn,xn), exactly one neutron (1n,1n),
or no neutrons (0n,0n) in each ZDC; the latter occurs only in the
topology trigger.

Figure~\ref{fig:all}d) shows the $d\sigma/dM_{\pi\pi}$ spectrum for
events with pair-$p_T\!< \!150$~MeV/c (points).  The fit (solid) is
the sum of a relativistic Breit-Wigner for $\rho^0$ production and a
S\"oding interference term for direct $\pi^+\pi^-$
production~\cite{soeding} (both dashed). A second order polynomial
(dash-dotted) describes the combinatorial background (shaded
histogram) from grazing nuclear collisions and incoherent
photon-nucleon interactions.  Incoherent $\rho^0$ production, where a
photon interacts with a single nucleon, yields high $p_T$ $\rho^0$s,
which are suppressed by the low pair $p_T$ requirement; the remaining
small contribution is indistinguishable from the coherent process.  A
coherently produced background arises from the two-photon process
$AuAu\! \rightarrow\! Au^{(\star)} Au^{(\star)} l^+l^-$. It
contributes mainly at low invariant mass $M_{\pi\pi} \!<\! 
0.5$~GeV/c$^2$. The small contribution from $\omega$ decays is
neglected.

\begin{wrapfigure}[11]{r}{.5\textwidth}
\includegraphics[width=7.5cm,height=2.8cm,bbllx=0pt,bblly=30pt,bburx=520pt,bbury=260pt]{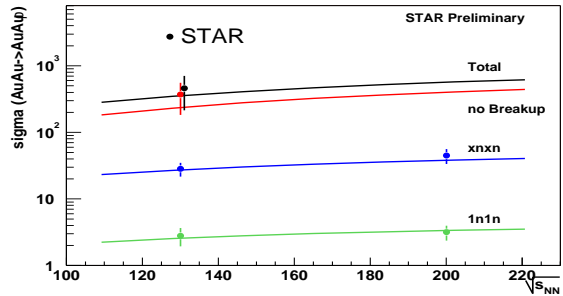}
\caption{Comparison to predictions from Ref.~\cite{BKN}. 
\label{fig:comp}}
\end{wrapfigure}

Figure~\ref{fig:comp} compares our results on cross sections for
coherent $\rho^0$ production at $\sqrt{s_{NN}}=130$~GeV\cite{starrho}
to the calculations of Ref.~\cite{BKN}. Preliminary results for
$\sqrt{s_{NN}}=200$~GeV are also shown in the plot.  The cross
sections are obtained from the integral of the Breit-Wigner fit,
extrapolated to full rapidity.  The integrated luminosity for the
minimum bias data is determined from the number of hadronic
interactions, assuming a total gold-gold hadronic cross section of
$7.2$~b~\cite{xsectAuAu}.  For coherent $\rho^0$ production
accompanied by mutual nuclear break-up (xn,xn), we measure a cross
section of \xsxnxn.  By selecting single neutron signals in both ZDCs,
we obtain \xssnsn.  The systematic uncertainties are  dominated by
the uncertainties of the luminosity determination and the $4\pi$
extrapolation.  The absolute efficiency of the  topology
trigger is poorly known and does not allow a direct cross section
measurement.  From $\sigma(AuAu\!\rightarrow\! Au^*_{xn} Au^*_{xn}
\rho^0)$ and the ratio $\sigma^\rho_{xn,xn}/\sigma^\rho_{0n,0n}$ we
estimate \xsnobrk~  and the total cross section for coherent $\rho^0$ production \xstot.

\begin{wrapfigure}[15]{r}{.6\textwidth}
\begin{minipage}[t]{4.6cm}
\includegraphics[width=4.8cm,height=3.2cm,bbllx=0pt,bblly=30pt,bburx=570pt,bbury=400pt]{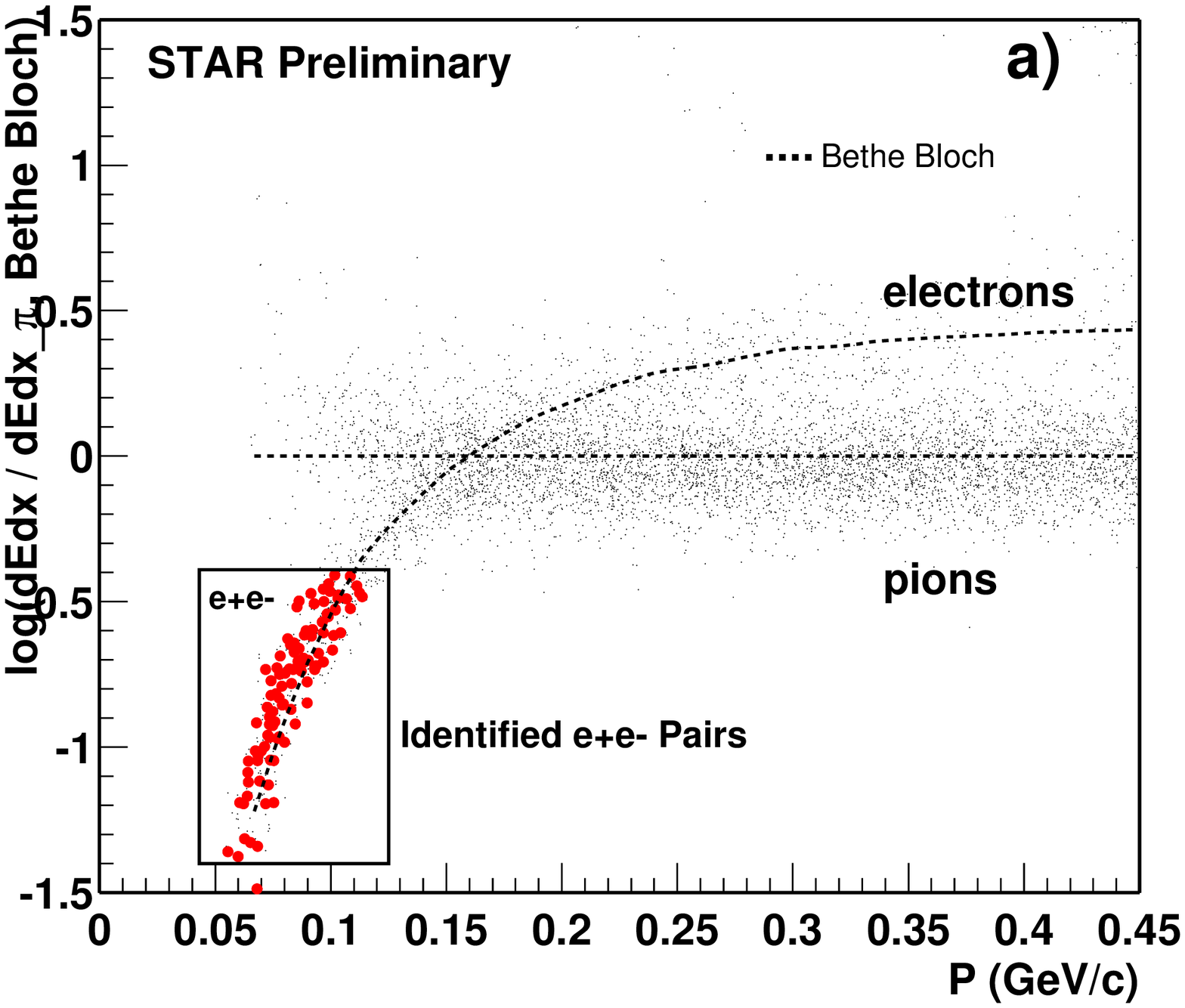}
\end{minipage}
\hspace{\fill}
\begin{minipage}[t]{4.6cm}
\includegraphics[width=4.8cm,height=3.2cm,bbllx=0pt,bblly=30pt,bburx=570pt,bbury=400pt]{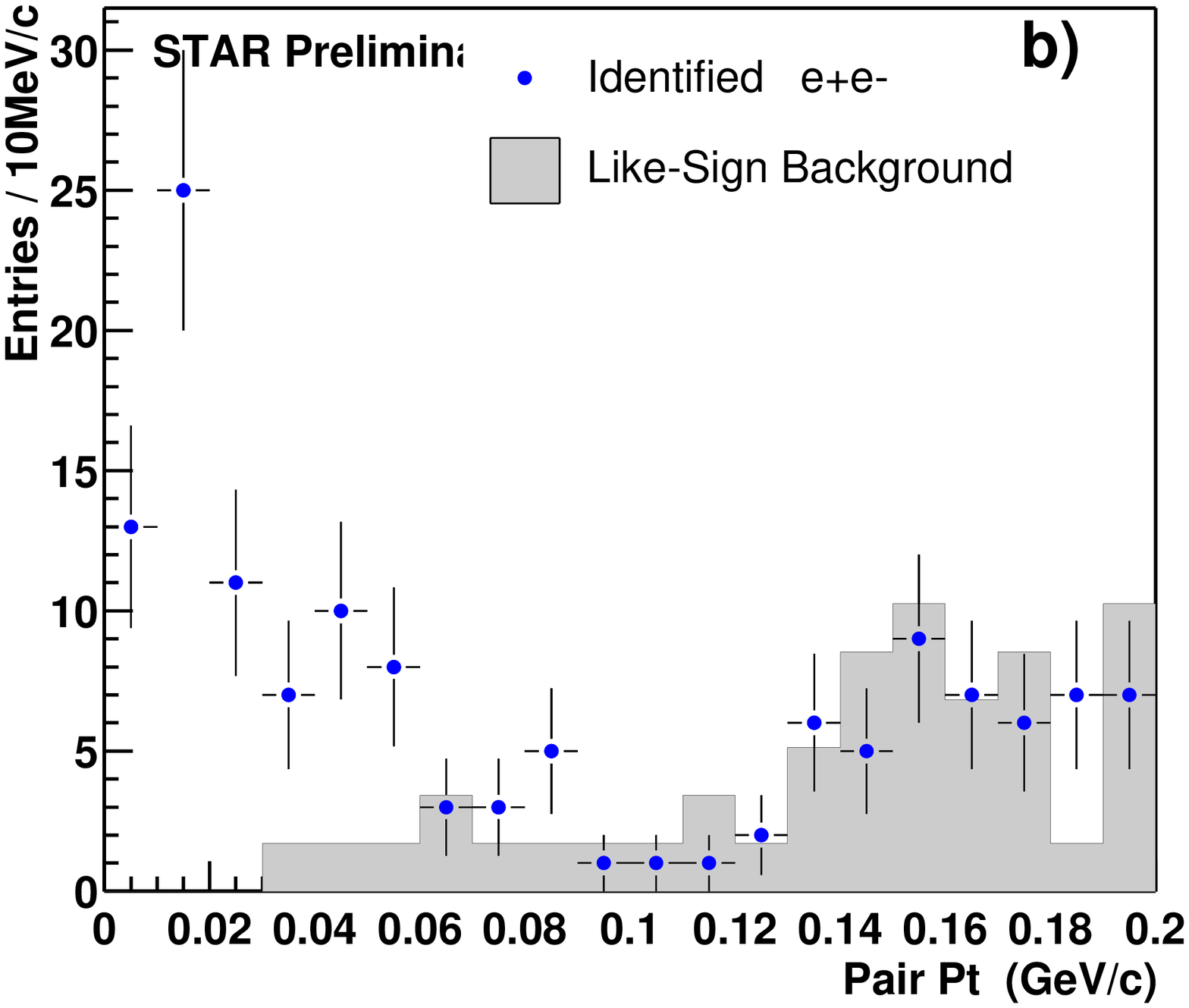}
\end{minipage}
\caption{(a) Energy loss $dE/dx$ of tracks in the 2-track, $\sqrt{s_{NN}}=200$~GeV minimum bias data
with reduced magnetic field. The dots indicate
events where both particles are identified as electrons. (b) The $p_T$
spectrum for identified $e^+e^-$ pairs. \label{fig:electrons} }
\end{wrapfigure}
Two-photon interactions include the purely electromagnetic process of
$e^+e^-$ pair production as well as single and multiple meson
production.  The coupling $Z\alpha$ ($0.6$ for $Au$) is large, so
$e^+e^-$ pair production is an important probe of quantum
electrodynamics in strong fields~\cite{baurrev}.  At momenta below
$125$ MeV, $e^+e^-$ pairs are identified by their energy loss in the
TPC as shown in Fig.~\ref{fig:electrons}a) for a $\sqrt{s_{NN}}=200$~GeV minimum bias data
sample. This data was taken with a $0.25$~T magnetic 
field;  at the full $0.5$~T field the low momentum electrons are bent out of the 
acceptance. Figure~\ref{fig:electrons}b shows the $p_T$ spectrum
for identified $e^+e^-$ pairs; a clear peak at $p_T \sim 1/b \!< \!20$
MeV/c identifies the process $AuAu \rightarrow Au^\star Au^\star
e^+e^-$.

In summary, ultra-peripheral heavy-ion collisions are a new
laboratory for diffractive interactions, complementary to fixed-target
$\rho^0$ photo-production on complex nuclei~\cite{alvensleben}.
The first measurements of coherent $\rho^0$ production
with and without accompanying nuclear excitation, $AuAu
\!\rightarrow\! Au^\star Au^\star \rho^0$ and $AuAu \!\rightarrow\! Au
Au \rho^0$, confirm the existence of vector meson production in
ultra-peripheral heavy ion collisions.
The cross sections at $\sqrt{s_{NN}}\!=\!130$ and $200$~GeV are in
agreement with theoretical calculations.

\vspace*{-1cm}
\end{document}